\documentclass[doublecolumn]{article}
\usepackage{spconf}

\usepackage{amsmath,amssymb,graphicx}
\usepackage[colorlinks=true,bookmarks=false,citecolor=blue,urlcolor=blue]{hyperref} 

\usepackage{graphicx}
\usepackage[english]{babel}
\usepackage[T1]{fontenc}  
\usepackage[latin1]{inputenc} 
\usepackage{url}
\usepackage{graphicx}
\usepackage{subfigure}
\usepackage{xcolor}
\usepackage{amsmath,amsfonts,amssymb}
\usepackage{booktabs}
\usepackage{epsfig}
\usepackage{pstricks}
\usepackage{pst-node}
\usepackage{pst-grad}
\usepackage{ifthen}
\usepackage{algorithm}
\usepackage{algorithmic}	
\usepackage{setspace}           

\usepackage{epstopdf}	
\usepackage{enumerate} 
\usepackage{hyperref}
\usepackage{multicol} 
\usepackage{enumitem} 
\usepackage{cite} 

\usepackage{verbatim} 
\usepackage{siunitx} 

\usepackage{fancyhdr}

\allowdisplaybreaks

\newcommand{\mypar}[1]{\bigskip\noindent {\bf #1}}

\definecolor{red}{RGB}{153,0,0}

\newcommand{\bX}{\mathbf{X}}

\newcommand{\bZ}{\mathbf{Z}}

\newcommand{\bD}{\mathbf{D}}

\newcommand{\bS}{\mathbf{S}}

\newcommand{\bz}{\mathbf{z}}

\newcommand{\bd}{\mathbf{d}}

\newcommand{\bW}{\mathbf{W}}

\title{Model-inspired Deep Learning for Light-Field Microscopy with Application to Neuron Localization}

\name{
	{Pingfan~Song$^{\star}$}, 
	{Herman V. Jadan$^{\star}$},
	{Carmel Howe$^{\dagger\ddag}$},
	{Peter Quicke$^{\dagger\ddag}$},
	{Amanda Foust$^{\dagger\ddag}$},
	{Pier Luigi Dragotti$^{\star}$}
	\thanks{This work was supported by the Biotechnology and Biological Sciences Research Council (BBSRC grant: BB/R009007/1), Wellcome Trust Seed Award (201964/Z/16/Z), Royal Academy of Engineering Research Fellowship (RF1415/14/26), Engineering and Physical Sciences Research Council (EPSRC grant:EP/L016737/1).}
}
\address{
	$^{\star}$ Department of Electronic and Electrical Engineering,
	$^{\dagger}$ Department of Bioengineering, 
	\\ 
	$^{\ddag}$ Center for Neurotechnology, Imperial College London, London, UK 
}

\begin{document}

\maketitle

\begin{abstract}
	%
	Light-field microscopes are able to capture spatial and angular information of incident light rays. This allows reconstructing 3D locations of neurons from a single snap-shot.
	%
	In this work, we propose a model-inspired deep learning approach to perform fast and robust 3D localization of sources using light-field microscopy images. This is achieved by developing a deep network that efficiently solves a convolutional sparse coding (CSC) problem to map Epipolar Plane Images (EPI) to corresponding sparse codes. The network architecture is designed systematically by unrolling the convolutional Iterative Shrinkage and Thresholding Algorithm (ISTA) while the network parameters are learned from a training dataset. Such principled design enables the deep network to leverage both domain knowledge implied in the model, as well as new parameters learned from the data, thereby combining advantages of model-based and learning-based methods. Practical experiments on localization of mammalian neurons from light-fields show that the proposed approach simultaneously provides enhanced performance, interpretability and efficiency.
\end{abstract}




\section{Introduction}

\vspace{-0.3cm}




Studying fundamental structural and dynamic properties of neuronal networks critically depends on advanced optical microscopy imaging systems.
%
In contrast to conventional optical microscopy that records only lateral information as a 2D projection of light rays, light-field microscopy (LFM)~\cite{levoy2006light,broxton2013wave,cohen2014enhancing} has has emerged as a promising 3D optical imaging technique. LFM is able to simultaneously gather both position and angular information of the incident light rays arriving at the sensor with a single snapshot~\cite{levoy1996light}. This is achieved by inserting a microlens array (MLA) at the native imaging plane and by moving the imaging sensor to the back focal plane of the microlenses~\cite{levoy2006light}. 
With benefits of high light efficiency and fast imaging speed, LFM demonstrates great potential of observing structures and dynamics at the cellular resolution across whole brain volumes~\cite{nobauer2017video,pegard2016compressive,quicke2020subcellular}.


\begin{figure}[t]
	\centering
	\begin{minipage}[b]{1\linewidth}
		\centering
		\includegraphics[width = 8cm, 
		trim=7cm 0cm 0cm 0cm,clip
		]{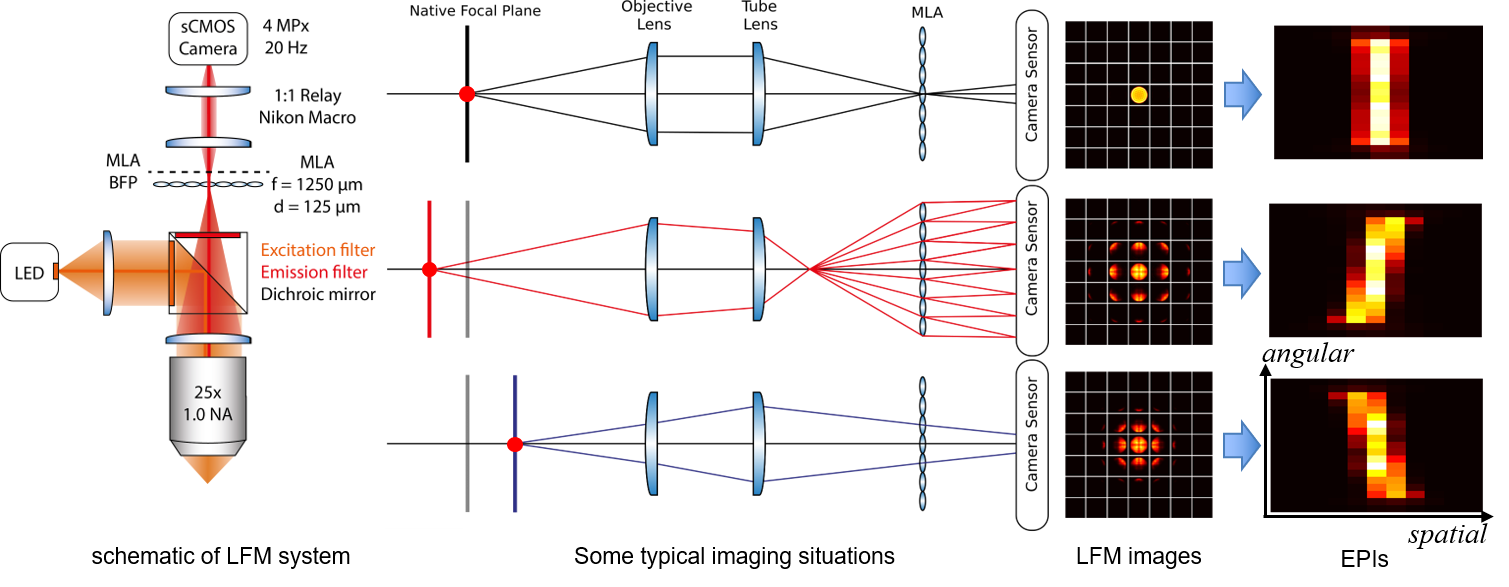}
	\end{minipage}
	
	\vspace{-0.3cm}
	
	\caption{Schematic for microlens-based LFM system and typical imaging situations, e.g. source in-focus and out-of-focus.
		Different depths (i.e. axial positions) cause different patterns in LFM images and constructed EPIs.
	}
	\label{Fig:microlens-basedLF}
\end{figure}

However, imaging with LFM also has limitations, such as reduced spatial resolution, slow reconstruction, image degradation due to light scattering in deep layers. 
To address these issues, some work focuses on improving spatial resolution via, e.g. 3D deconvolution~\cite{broxton2013wave,cohen2014enhancing,nobauer2017video}, while some work detects the 3D positions of sources directly by performing sparse decomposition in phase-space~\cite{liu20153d,pegard2016compressive}, or by performing convolutional sparse coding with respect to a synthetic dictionary~\cite{song20203D,song19location}. Different from existing methods, we propose a model-inspired deep learning approach for fast and robust 3D localization of sources from a single light-field image.

In the past decade, learning-based methods, in particular deep learning~\cite{goodfellow2016deep}, have demonstrated enhanced performance in terms of speed and accuracy over conventional model-based methods. However, deep neural networks are usually designed empirically and their structures lack interpretability, which is a prominent shortcoming. In the past few years, algorithm unrolling/unfolding~\cite{gregor2010learning,chen2018theoretical,li2020efficient,deng2020deep} has emerged as a promising technique to design deep networks in a more principled way by unrolling iterative methods. It bridges model-based methods with learning-based methods, and this leads to enhanced interpretability and better generalization ability of deep networks.

In this work, we develop a network using algorithm unrolling to localize neurons in tissues from LFM images. The network performs convolutional sparse coding on input epipolar plane images (EPI)~\cite{bolles1987epipolar,gortler1996lumigraph,vagharshakyan2018light}, a type of spatio-angular feature constructed from light-field images, and output sparse codes which indicates depth positions.




\begin{figure*}[t]
	\centering
	\includegraphics[width = 16cm, 
	]{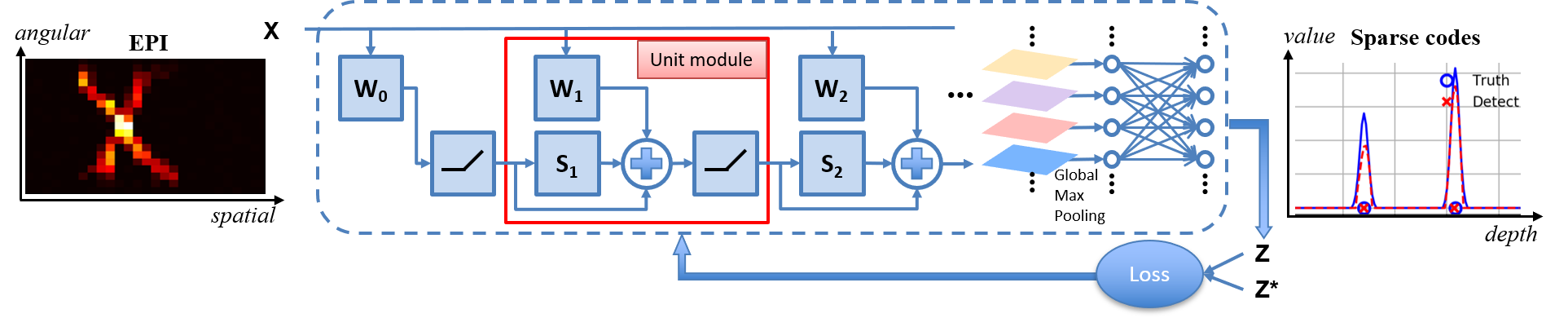}
	
	\vspace{-0.3cm}
	
	\caption{Proposed CISTA-net. An input EPI $\bX$ containing two sources is mapped to sparse codes $\bZ$ whose support indicates depths. Cross-entropy loss between the estimation $\bZ$ and the label $\bZ^\star$ is computed and back-propagated to update parameters.
	}
	\label{Fig:CISTA_net}
\end{figure*}

\vspace{-0.2cm}

\section{Model-inspired Network Design}
\label{sec:Problems}

\vspace{-0.2cm}


To design the network in a principled and systematic way, we first present the convolutional sparse coding (CSC) model~\cite{song20203D} and a convolutional version of Iterative Shrinkage and Thresholding Algorithm (ISTA)~\cite{beck2009fast}. Then, we unroll the Convolutional ISTA method to design the network architecture, so that the network, referred to as CISTA-net, imitates the processing flow that maps EPIs into sparse codes. Finally, the parameters of the network are learned from a well-designed training dataset. 


\vspace{-0.2cm}

\mypar{Convolutional Sparse Coding model}

The method developed in~\cite{song20203D} leverages the fact that neurons localized in space are relatively similar to point-like sources. Given raw light-field microscopy images, we use the method proposed in~\cite{song20203D} to perform calibration, conversion, and purification to get an array of clean sub-aperture images. Then, the horizontal and vertical positions, i.e. the x and y coordinates of sources can be detected from the central sub-aperture image by finding those pixels brighter than a specified threshold. 

Depth detection along the axial coordinate is not trivial as it requires the proper leverage of the angular information. To this end, EPIs are often used to simultaneously reveal the spatial and angular information captured in 4D light-field data, as shown in Figure~\ref{Fig:microlens-basedLF}. 
%
To infer depths from EPIs, one may construct an EPI dictionary that consists of basic EPIs related to sources which are associated to different depths.
Such an EPI dictionary can be used to decompose an input EPI into a sparse linear combination of basic EPIs. 
Finding this sparse decomposition can be modelled as a convolutional sparse coding problem~\eqref{Eq:CSC_model}(See also~\cite{song20203D}).

Specifically, given an observed EPI $\bX \in \mathbb{R}^{\Theta \times N}$ where $\Theta$ denotes angular dimension, i.e. number of pixels in a single microlens, and $N$ denotes spatial dimension, i.e. number of microlens, we aim to obtain a series of sparse codes $\{ \bz_m \in \mathbb{R}^{\Theta \times N} \}_{m=1}^M$ via decomposition with respect to a predefined EPI dictionary $\{ \bd_m \in \mathbb{R}^{\theta \times n} \}_{m=1}^M$ ($M$ denotes number depths to be covered and $\theta \leq \Theta$, $n \leq N$). For simplicity, all the variables have been vectorized into column vectors and then such decomposition can be formulated as:
\begin{equation}
\label{Eq:CSC_model}
\begin{array}{cl}
\underset{ \{\bz_m \}}{\min}
&
\frac{1}{2} \left\|\bX - \sum\limits_{m=1}^{M} \bd_m * \bz_m \right\|_2^2
+ \lambda \, \sum\limits_{m=1}^{M} \|\bz_m \|_1
\end{array}
\end{equation}
where, $\bd_m$ is the $m$-th element of the EPI dictionary and $\| \cdot \|_1$ is the $\ell_1$ norm that summarizes the absolute value of non-zero elements.

%

\mypar{Convolutional ISTA algorithm.}
Problem~\eqref{Eq:CSC_model} can be solved using the Convolutional ISTA algorithm which consists of the following computationally-efficient projected gradient descent iterations: 
\begin{equation} \label{Eq:conv_ISTA_SingleUnit}
\small
\begin{split}
\bz_k^{(i+1)} 
& = \mathcal{T}_{\lambda} \left( \bz_k^{(i)} - \gamma \bd_k^\top * \sum\limits_{m=1}^{M} \bd_m * \bz_m^{(i)} + \bd_k^\top * \bX \right)
\end{split}
\end{equation}
where $k = 1, \cdots, M$, the superscript $^{(i)}$ denotes the iteration index, and $\gamma$ is the step size.
$\mathcal{T}_{\lambda} (\cdot)$ is the element-wise soft-thresholding function, defined as $\mathcal{T}_\lambda(x) = \text{sign}(x) \cdot (| x | - \lambda)_+$ with an appropriate $\lambda \geq 0$, which is used to enforce sparsity on sparse codes.
Equation~\eqref{Eq:conv_ISTA_SingleUnit} can also be rewritten as:
\begin{equation} \label{Eq:conv_ISTA_Unit}
\small
\begin{split}
\bZ^{(i+1)} 
& = \mathcal{T}_{\lambda} \left( \bZ^{(i)} - \gamma \bD^\top \circledast 
\sum \bD \circledast \bZ^{(i)} 
+ \bD^\top \circledast \bX \right)
\end{split}
\end{equation}
where dictionary $\bD \in \mathbb{R}^{\theta \times n \times M}$ consists of an array of $M$ $\theta \times n$ basic EPIs, $\bD^\top \in \mathbb{R}^{n \times \theta \times M}$ is the transposed version of $\bD$. Symbol $\circledast$ denotes convolution along the first and second dimension, i.e. spatio-angular dimension, and the symbol $\sum$ denotes summation of convolution results along the third dimension.
Once the sparse codes are obtained, the largest values of each sparse code $\bz_m$ are selected and combined into a new vector, similar to the GlobalMaxPooling operation commonly used in deep learning. Then the non-zero support of the new sparse vector leads to the depth.

\mypar{CISTA-net design.}
Our CISTA-net aims to mimic above processing flow. So \eqref{Eq:conv_ISTA_Unit} is adapted to \eqref{Eq:CISTA_net_block} which serves as the unit module to build the network. Then, the entire network architecture can be obtained by concatenating the unit module multiple times, which accounts for algorithm unrolling. So we have:
\begin{equation} \label{Eq:CISTA_net_block}
\small
\begin{split}
\bZ^{(i+1)} 
& = ReLU \left( \bZ^{(i)} - \bS^{(i)} \circledast \bZ^{(i)} + \bW^{(i)} \circledast \bX + \lambda \right)
\end{split}
\end{equation}
where trainable filters $\bS^{(i)}$ and $\bW^{(i)}$ replace the variable $\gamma \bD^\top \circledast \sum \bD $ and $\gamma \bD^\top$, respectively. $ReLU$ (i.e. Rectified Linear Unit) with trainable non-negative bias $\lambda$ play the role of the soft-thresholding. The overall architecture of the CISTA-net is shown in Figure~\ref{Fig:CISTA_net}.

Even though the unrolling/unfolding idea has been investigated in the literature, the proposed unrolling network has some particular characteristics and appealing attributes. In particular, operation $ \bd_m * \bz_m^{(i)}$ in \eqref{Eq:conv_ISTA_SingleUnit} or $ \bD \circledast \bZ^{(i)} $ in~\eqref{Eq:conv_ISTA_Unit} actually performs convolution between each pair of EPI $\bd_m$ and sparse codes $\bz_m$. Such pair-wise convolution inspires us to exploit, instead of plain convolution, depth-wise separable convolution~\cite{howard2017mobilenets} which reduces the number of network parameters. 
Besides, the subtraction operation $\bZ^{(i)} - \bS^{(i)} \circledast \bZ^{(i)}$ leads to a local skip connection which evokes the residual module used in the ResNet~\cite{he2016deep}. Finally, transformed $\bX$ added in each layer facilitates information propagation from the first layer directly to each hidden layer, thereby alleviating information loss. This structure reveals certain resemblance to dense connectivity module used in DenseNet~\cite{huang2017densely}, even though not each layer is receiving a connection from all preceding layers.
Even though some of these architecture modules have been used in modern neural networks, we here provide a new and interesting perspective to elaborate how they can naturally derive from a well-designed model and corresponding iterative method, rather than pure intuition commonly used in general network design.



Rather than sticking to Model~\eqref{Eq:conv_ISTA_Unit} rigidly, we also employ some customized modifications to further enhance the capability of the network. For example, the size of filters $\bS^{(i)}$ and $\bW^{(i)}$ increases across different layers so that the receptive field increases gradually to facilitate capture of both local details and global structure.
In addition, a fully-connected layer followed by sigmoid activation is added after the GlobalMaxPooling layer to perform non-linear transformation. Such slight departures from the original convolutional ISTA algorithm enable extended representation capability of the network.




%

\vspace{-0.2cm}

\section{Experiment}
\label{sec:Experiments}

\vspace{-0.2cm}

\begin{figure}[t]
	\begin{minipage}[b]{0.33\linewidth}
		\raggedleft
		\includegraphics[width = 2.4cm, 
		]{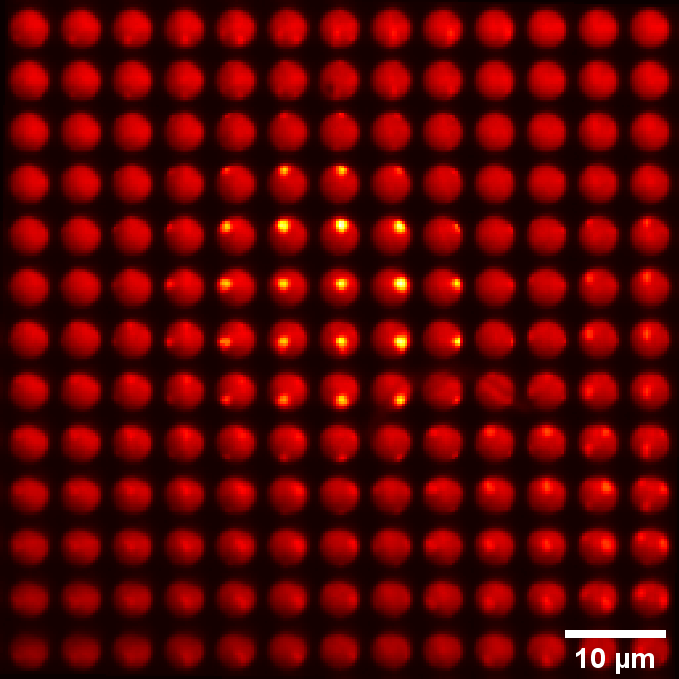}
	\end{minipage} 
	\begin{minipage}[b]{0.325\linewidth}
		\raggedleft
		\includegraphics[width = 2.4cm, 
		]{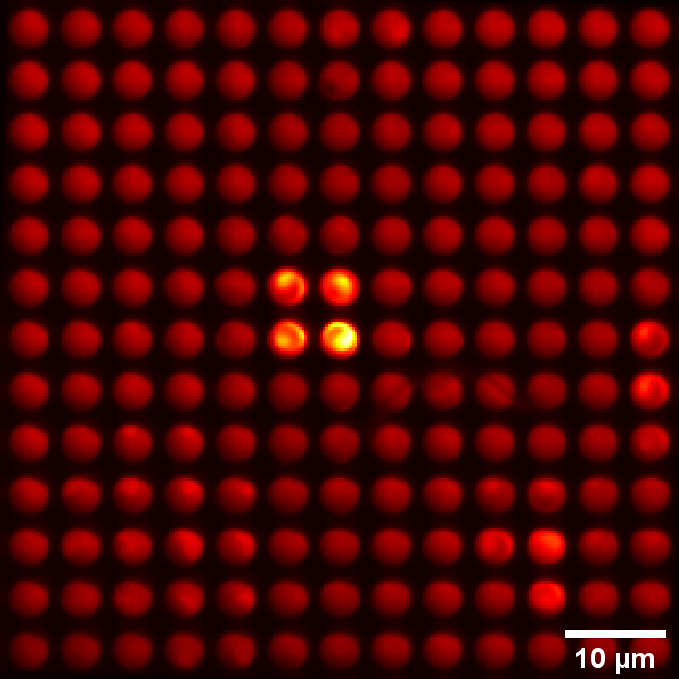}
	\end{minipage}
	\begin{minipage}[b]{0.325\linewidth}
		\raggedleft
		\includegraphics[width = 2.4cm, 
		]{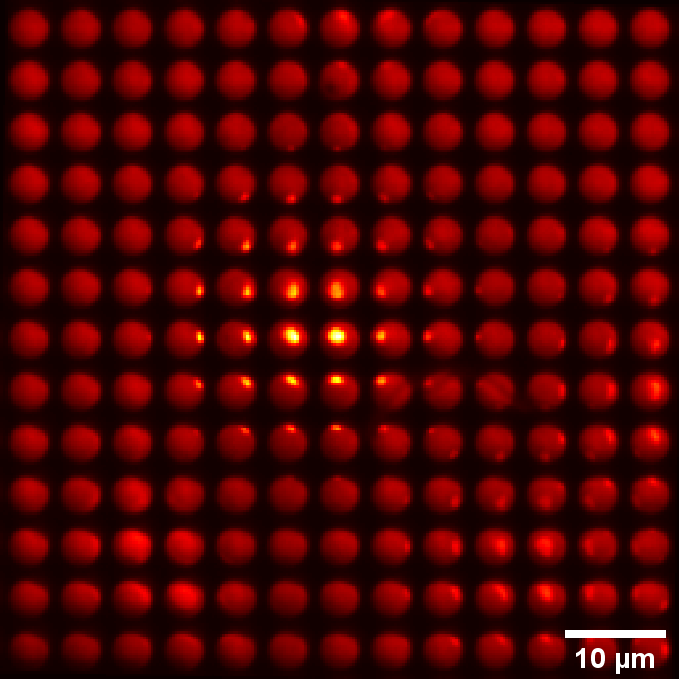}
	\end{minipage}
	\\
	\begin{minipage}[b]{1\linewidth}
		\raggedleft
		\includegraphics[width = 2.8cm, 
		trim=0cm 0cm 9cm 0.5cm,clip
		]{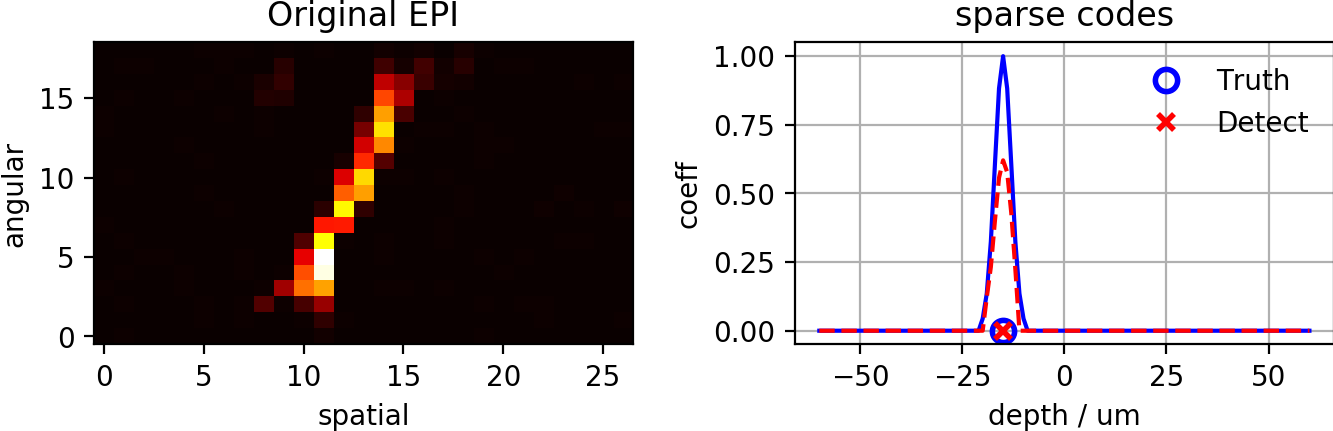}
		\includegraphics[width = 2.8cm, 
		trim=0cm 0cm 9cm 0.5cm,clip
		]{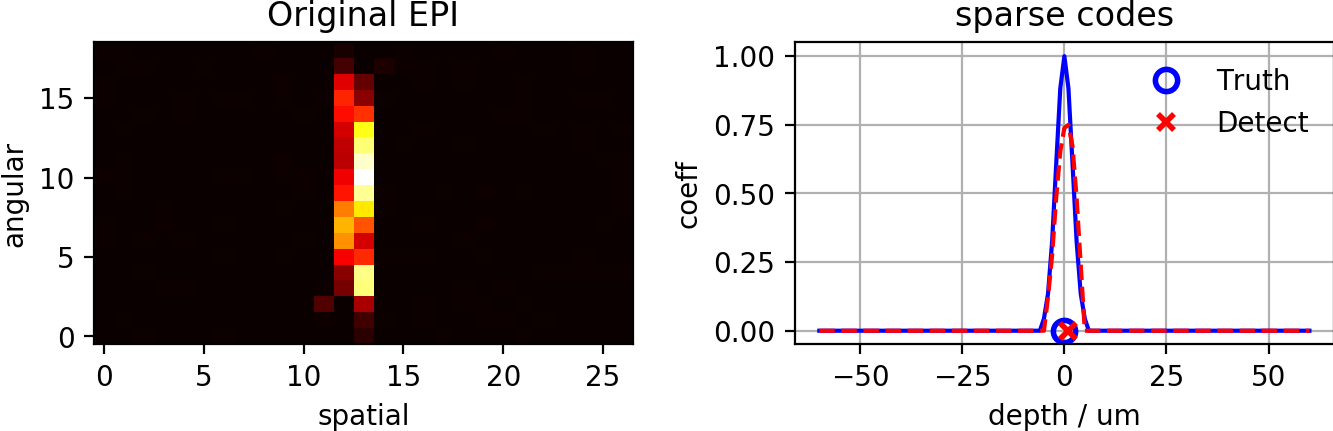}
		\includegraphics[width = 2.8cm, 
		trim=0cm 0cm 9cm 0.5cm,clip
		]{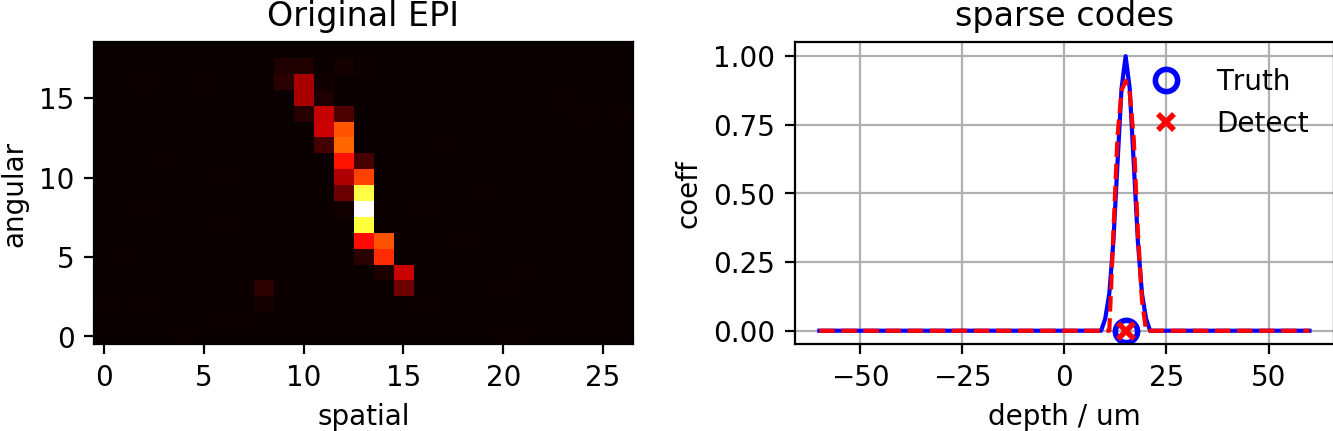}
		\\
		\scriptsize (a) LFM images (above) and corresponding EPIs (bottom) for depth of -15, 0, 15 \si{um}.
	\end{minipage} 
	\\
	\vspace{+0.2cm}
	\begin{minipage}[b]{1\linewidth}
		\centering
		\includegraphics[width = 2.8cm, 
		trim=0cm 0cm 0.8cm 0cm,clip
		]{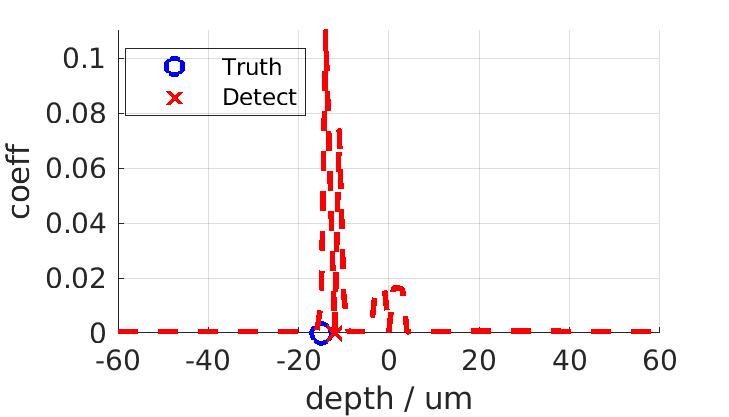}
		\includegraphics[width = 2.8cm, 
		trim=0cm 0cm 0.8cm 0cm,clip
		]{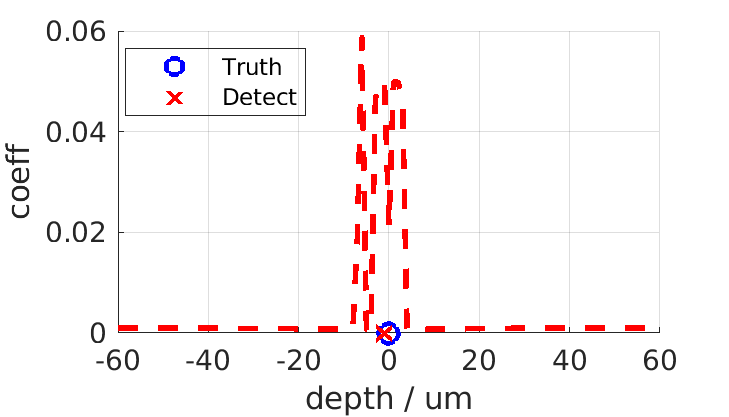}
		\includegraphics[width = 2.8cm, 
		trim=0cm 0cm 0.8cm 0cm,clip
		]{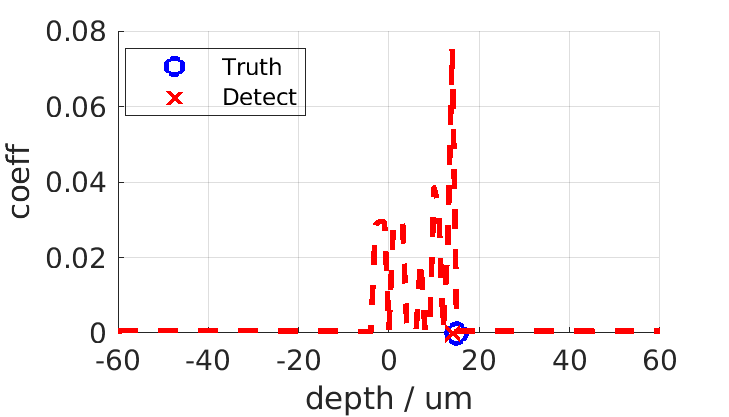}
		\\
		\scriptsize (b) Estimation of sparse codes and depth using phase-space method~\cite{liu20153d,pegard2016compressive}.
	\end{minipage} 
	\\
	\vspace{+0.2cm}
	\begin{minipage}[b]{1\linewidth}
		\centering
		\includegraphics[width = 2.8cm, 
		trim=0cm 0cm 0.8cm 0cm,clip
		]{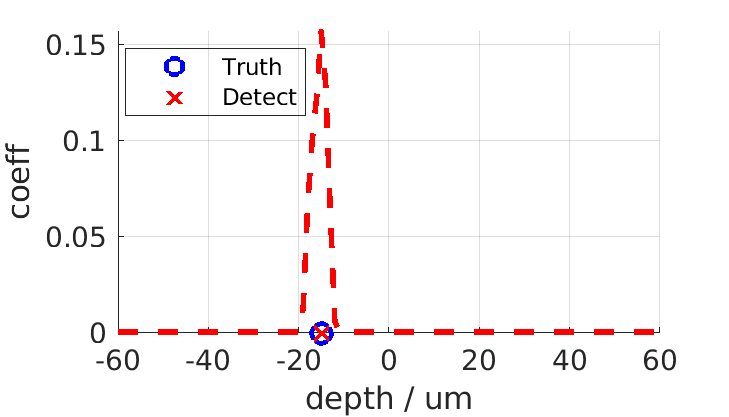}
		\includegraphics[width = 2.8cm, 
		trim=0cm 0cm 0.8cm 0cm,clip
		]{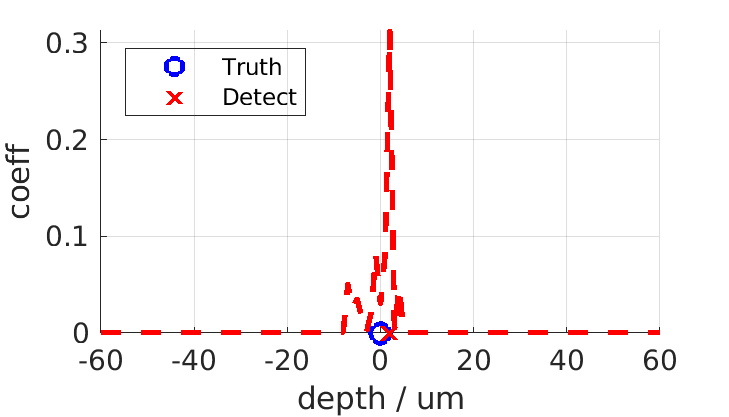}
		\includegraphics[width = 2.8cm, 
		trim=0cm 0cm 0.8cm 0cm,clip
		]{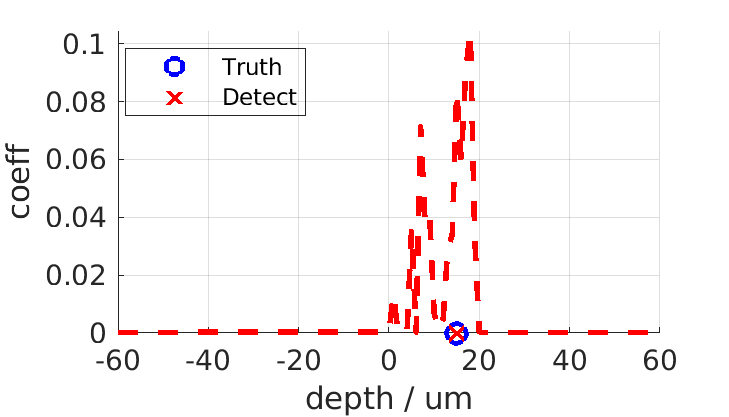}
		\\
		\scriptsize (c) Estimation of sparse codes and depth using CSC method~\cite{song20203D}.
	\end{minipage} 
	\\
	\vspace{+0.2cm}
	\begin{minipage}[b]{1\linewidth}
		\centering
		\includegraphics[width = 2.8cm, 
		trim=8.5cm 0cm 0cm 0.5cm,clip
		]{./Figures/CISTA-net/Cell_SZ257/noBG_55Frames_s1/EPI_Supp_Val_3.png}
		\includegraphics[width = 2.8cm, 
		trim=8.5cm 0cm 0cm 0.5cm,clip
		]{./Figures/CISTA-net/Cell_SZ257/noBG_55Frames_s1/EPI_Supp_Val_18.png}
		\includegraphics[width = 2.8cm, 
		trim=8.5cm 0cm 0cm 0.5cm,clip
		]{./Figures/CISTA-net/Cell_SZ257/noBG_55Frames_s1/EPI_Supp_Val_33.png}
		\\
		\scriptsize (d) Estimation of sparse codes and depth using CISTA-net.
	\end{minipage} 
	
	\vspace{-0.5cm}	
	
	\caption{
		Visual comparison of estimated sparse codes and depth detection using different methods. In each subfigure, the red dashed line indicates estimated sparse codes and the red crosses indicate estimated depths. The blue line indicates the soft-labels, and the blue circle indicates the groundtruth depths.
	}
	\label{Fig:SparseCodes}
\end{figure}

In this section, we evaluate the performance of CISTA-net on the 3D localization task. We also compare our approach with the phase-space based method (Phase-Space for short)~\cite{liu20153d,pegard2016compressive} and convolutional sparse coding based method (CSC for short)~\cite{song20203D} on light-field microscopy data obtained from scattering specimens -- genetically encoded fluorophore in mouse brain tissues, as shown in Fig.~\ref{Fig:SparseCodes} (a).

The raw light-field microscopy images were captured by systematically changing the axial distance between the specimens and the objective lens of LFM as in~\cite{song20203D}. Therefore, each light-field microscopy image captures a 3D volume at a specified depth. 
All the experiments were conducted in a computer equipped with an Intel hexa-core i7-8700U CPU@3.20GHz with 28GB of memory and a NVIDIA GTX 1080 Ti GPU.

\begin{figure}[t]
	\centering
	\begin{minipage}[b]{1\linewidth}
		\centering
		\includegraphics[width = 9cm, 
		]{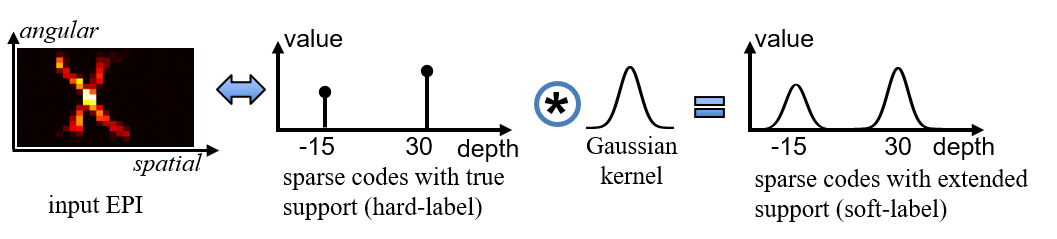}
	\end{minipage} 
	
	\vspace{-0.4cm}
	
	\caption{
		Illustration to the construction of soft-labels.
	}
	\label{Fig:Soft-label}
\end{figure}

\begin{figure}[t]
	\centering
	\begin{minipage}[b]{1\linewidth}
		\centering
		\includegraphics[width = 6cm, height = 3.5cm,
		]{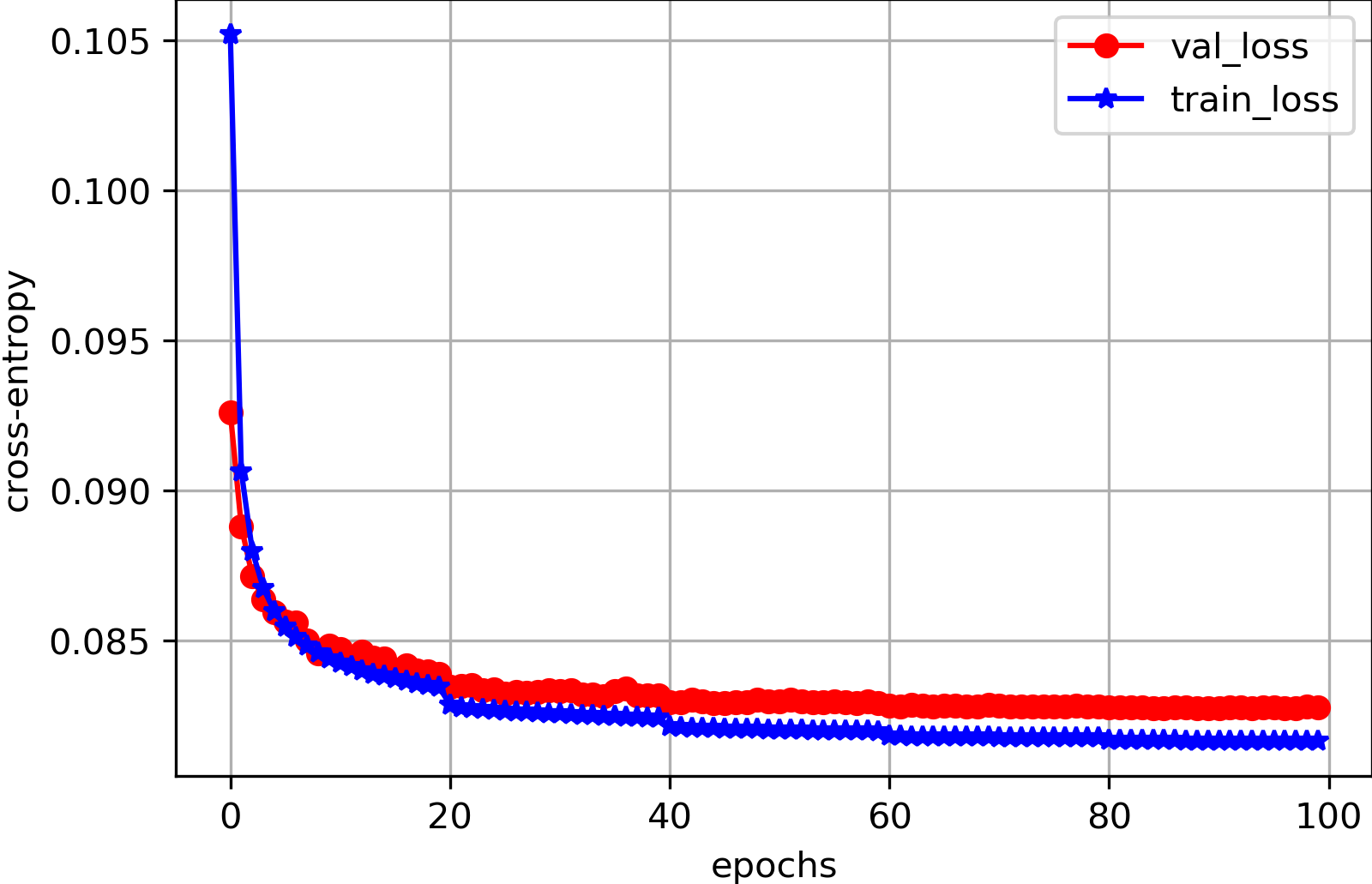}
	\end{minipage} 
	
	\vspace{-0.3cm}
	
	\caption{
		Convergence curve of training and validation loss.
	}
	\label{Fig:Training}
\end{figure}

\begin{figure}[thb]
	\centering
	\begin{minipage}[b]{0.32\linewidth}
		\centering 
		\includegraphics[width = 3cm, height = 2.8cm,
		]{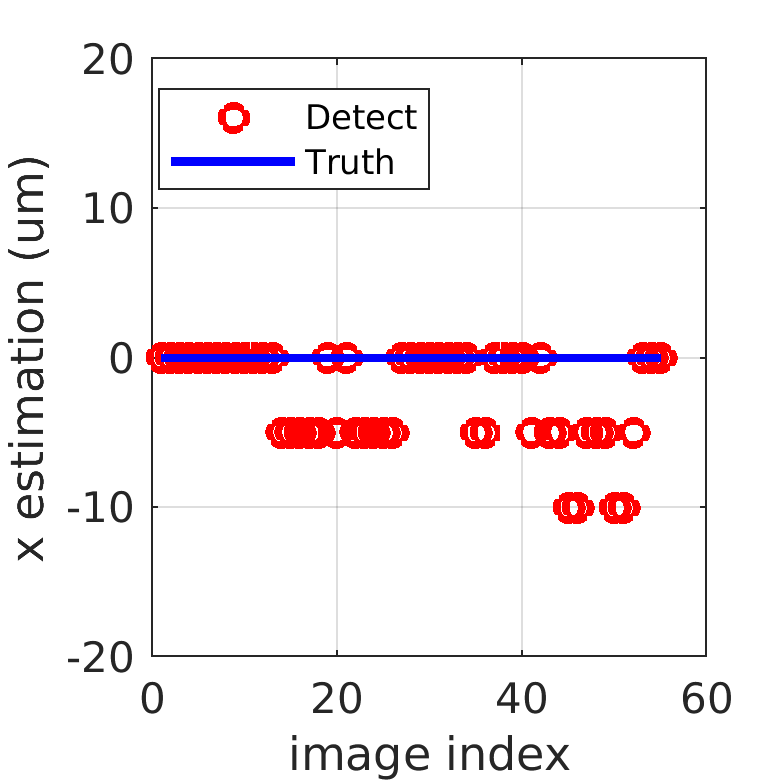}
	\end{minipage} 	
	\begin{minipage}[b]{0.32\linewidth}
		\centering 
		\includegraphics[width = 3cm, height = 2.8cm,
		]{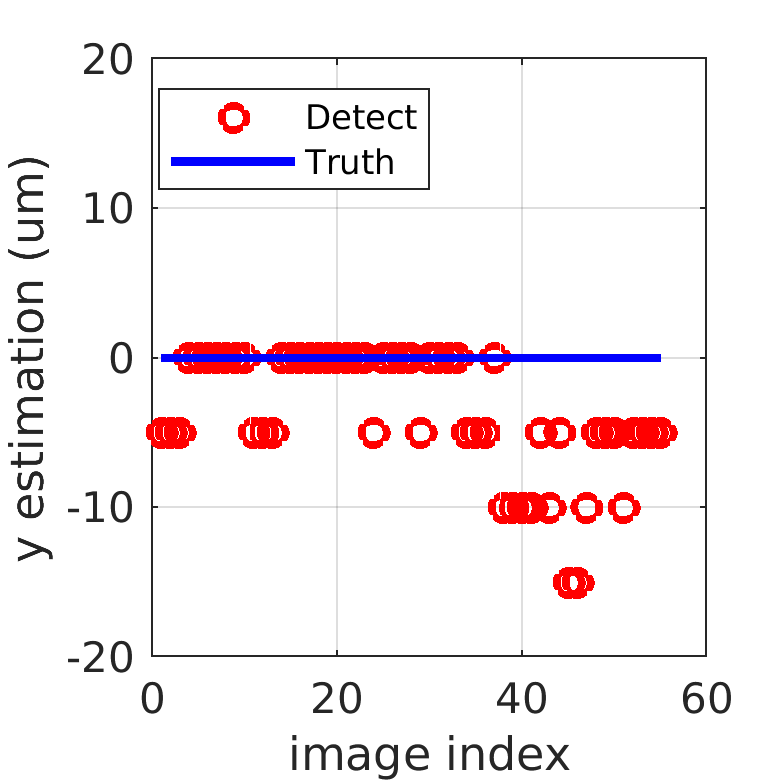}
	\end{minipage}
	\begin{minipage}[b]{0.32\linewidth}
		\centering 
		\includegraphics[width = 3cm, height = 2.8cm,
		trim=0.6cm 0cm 8.4cm 0cm,clip
		]{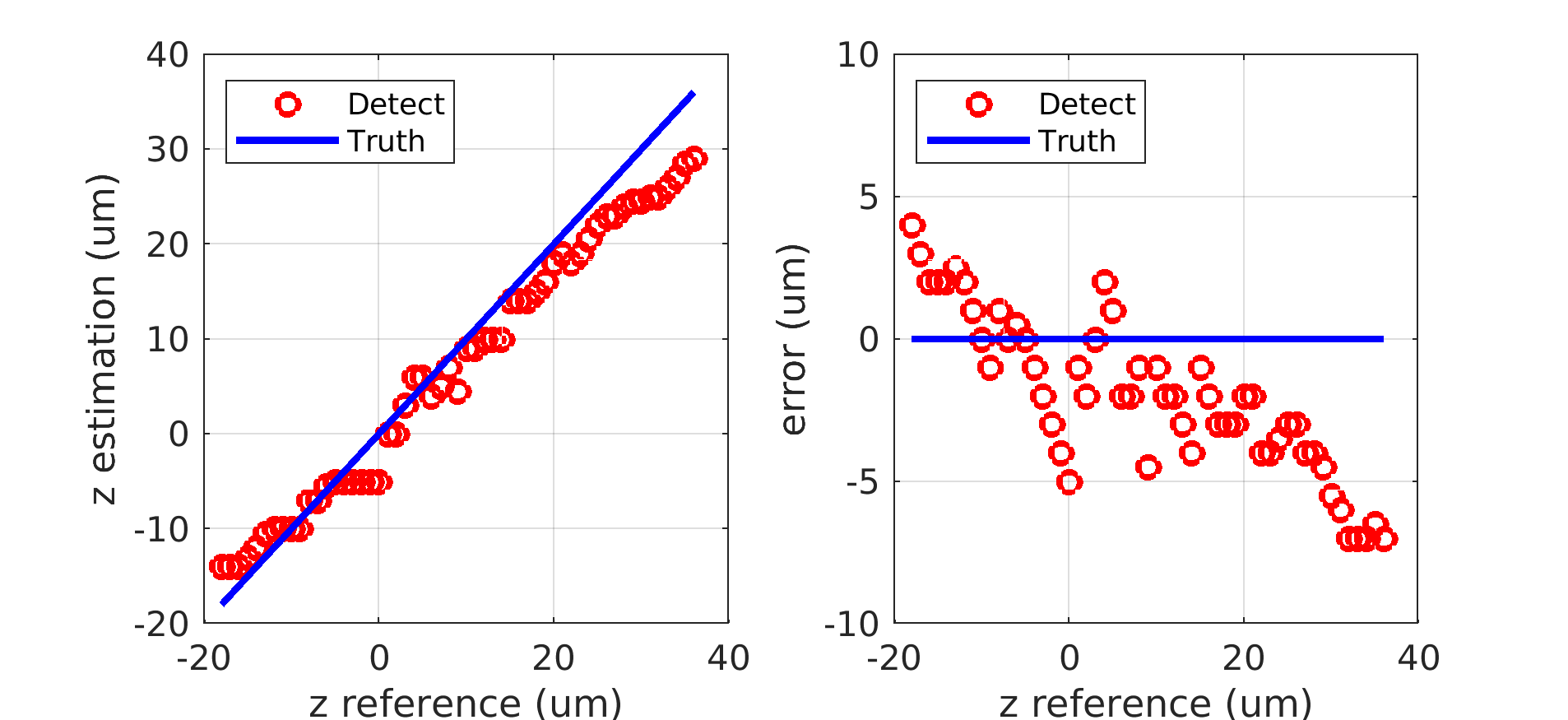}
	\end{minipage} 	
	\\
	\vspace{+0.2cm}
	\begin{minipage}[b]{0.99\linewidth}
		\centering 
		\small (a) Localization performance of phase-space method~\cite{liu20153d,pegard2016compressive}. RMSE for x, y, z position detection is 4.05, 5.48, 3.41 \si{\um}, respectively.
	\end{minipage} 	
	\\
	\vspace{+0.2cm}
	\begin{minipage}[b]{0.32\linewidth}
		\centering 
		\includegraphics[width = 3cm, height = 2.8cm,
		]{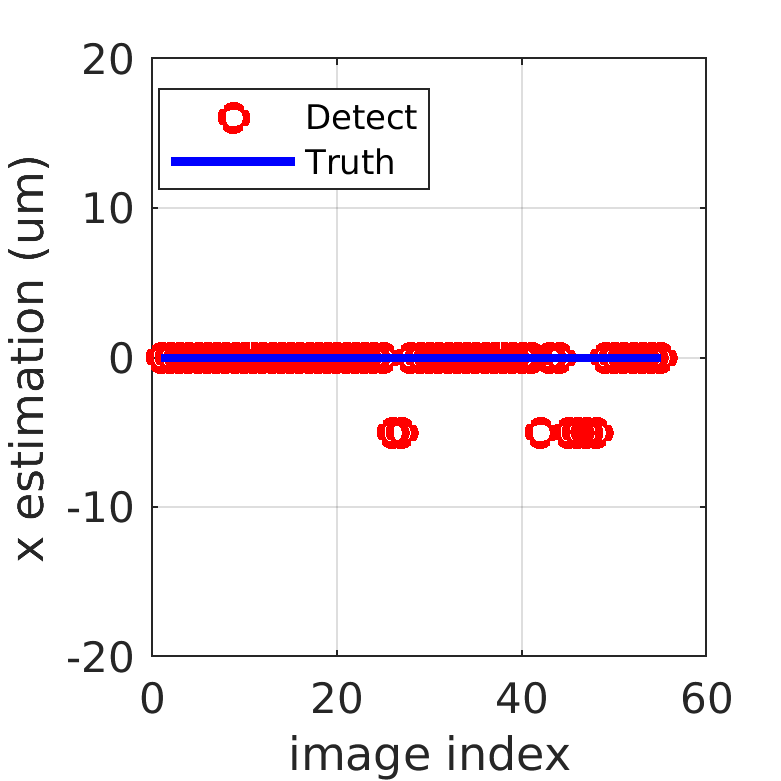}
	\end{minipage} 	
	\begin{minipage}[b]{0.32\linewidth}
		\centering 
		\includegraphics[width = 3cm, height = 2.8cm,
		]{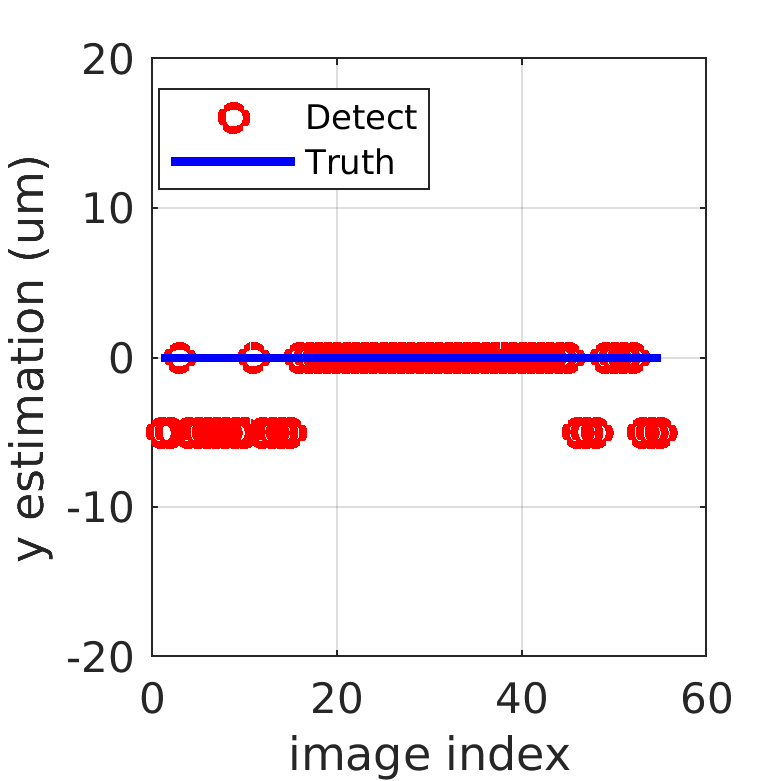}
	\end{minipage}
	\begin{minipage}[b]{0.32\linewidth}
		\centering 
		\includegraphics[width = 3cm, height = 2.8cm,
		trim=0.6cm 0cm 8.3cm 0cm,clip
		]{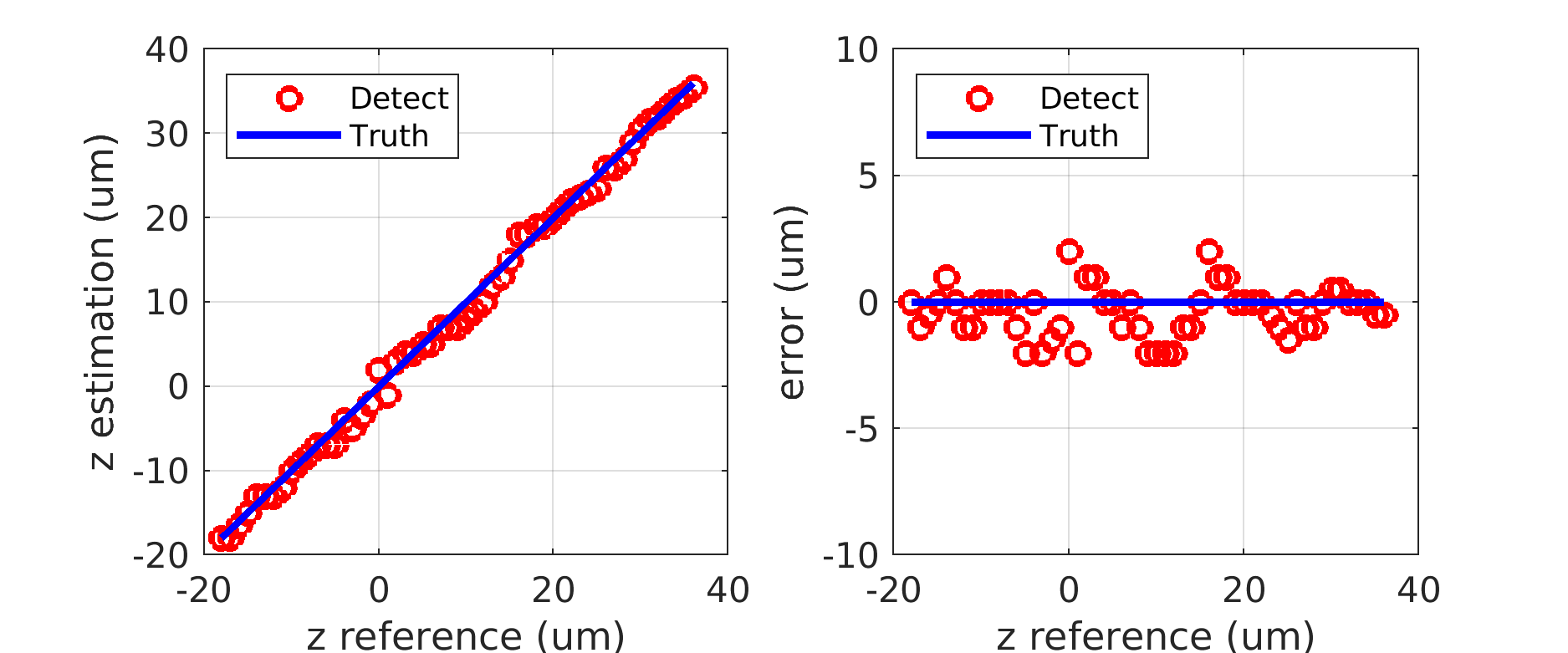}
	\end{minipage} 	
	\\
	\vspace{+0.2cm}
	\begin{minipage}[b]{0.99\linewidth}
		\centering 
		\small (b) Localization performance of CSC approach~\cite{song20203D}. RMSE for x, y, z position detection is 1.78, 2.94, 1.14 \si{\um}, respectively.
	\end{minipage} 	
	\\
	\vspace{+0.3cm}
	\begin{minipage}[b]{0.66\linewidth}
		\centering 
		\includegraphics[width = 5.6cm, height = 2.6cm,
		]{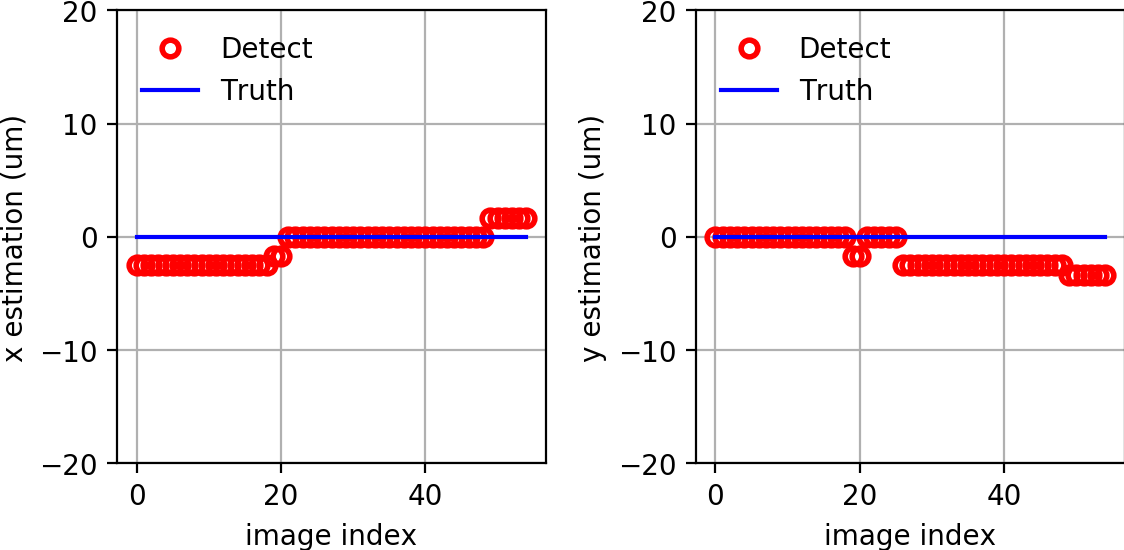}
	\end{minipage} 	
	\begin{minipage}[b]{0.32\linewidth}
		\centering 
		\includegraphics[width = 2.7cm, height = 2.6cm,
		trim=0cm 0cm 7.9cm 0cm,clip
		]{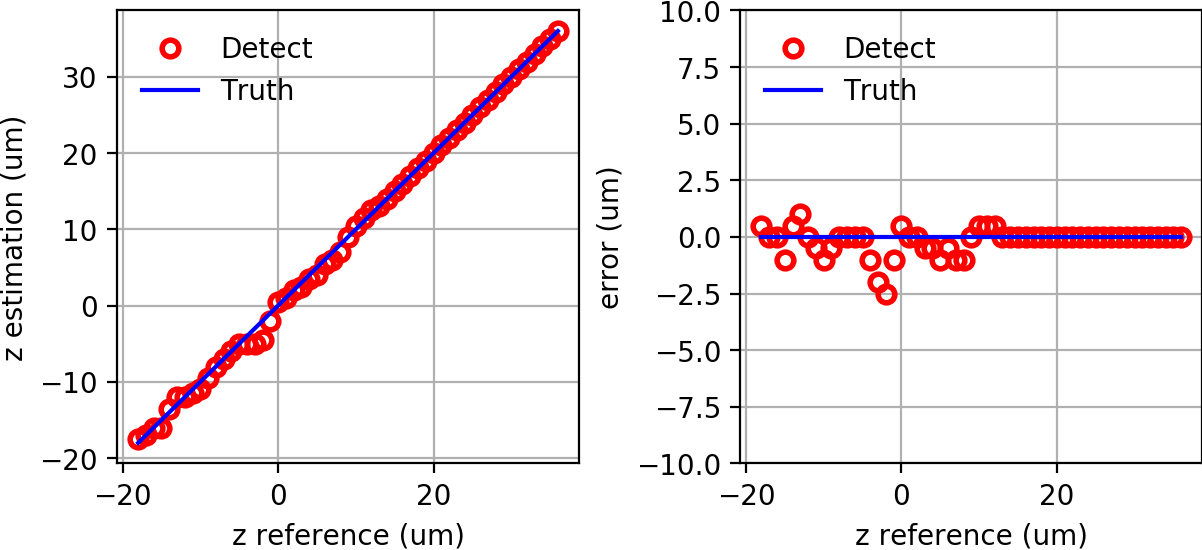} 
	\end{minipage} 	
	\\
	\vspace{+0.2cm}
	\begin{minipage}[b]{0.99\linewidth}
		\centering 
		\small (c) Localization performance of the proposed CISTA-net. RMSE for x, y, z position detection is 1.60, 1.98, 0.82 \si{\um}, respectively.
	\end{minipage} 	
	\\
	
	\vspace{-0.2cm}
	
	\caption{
		Performance of localizing neurons using three different methods, including phase-space method~\cite{liu20153d,pegard2016compressive}, CSC method~\cite{song20203D}, and the proposed method. The depth of neurons varies from -18 \si{\um} to 36 \si{\um}. 
	}
	\label{Fig:Compare3Methods_Cell}
\end{figure}

\vspace{-0.2cm}

\mypar{Training settings.}
Figure~\ref{Fig:Soft-label} shows how we construct the labelled training datasets. Since the designed CISTA-net is expected to output sparse codes where the positions of non-zero elements (i.e. support) indicate the depths corresponding to input EPIs, it needs to be trained on labelled data in order to learn the mapping from an input EPI to the corresponding sparse codes. However, coefficient values in sparse codes are unknown.
To handle this issue, we treat the task as a multi-class, multi-label classification task where the support of the sparse codes indicates target classes/categories while the coefficient values indicate probability or confidence of the input signal falling into each class.%
\footnote{
	Note, if the task is treated as a regression task, the output of the network will be a number that denotes the continuous depth, thus it can not handle the case with multiple neurons in the region of interest as this gives multiple lines in an EPI.
}
In this way, we only need weakly annotated sparse codes with roughly estimated coefficient values.
%
In addition, we found that EPIs corresponding to adjacent depths tend to have similar patterns, and thus exhibit high coherence and lead to group sparsity in CSC.
Based on this observation, we modified the sparse codes by convolving them with a Gaussian kernel so that the groundtruth non-zero support is extended to neighbouring areas that cover adjacent depths. 
We call the support-extended sparse codes "soft-labels" in comparison with the sparse codes with exact support, namely "hard-labels". Soft-labels give some training benefits by incorporating data correlation as guidance information and enforce group sparsity for the output of network. Since our task is regarded as a multi-class, multi-label classification task, the loss function is set to be binary cross-entropy between output sparse codes and the soft-labels.



\vspace{-0.2cm}

\mypar{Training and testing results.}
We trained the network on 100,000 EPI samples with a maximum of 100 epochs using the adaptive moment estimation (ADAM) optimizer. Figure~\ref{Fig:Training} shows that the training and validation errors converge almost within 40 epochs.
Figure~\ref{Fig:SparseCodes} and \ref{Fig:Compare3Methods_Cell} compares the proposed CISTA-net with Phase-Space~\cite{liu20153d,pegard2016compressive} and CSC~\cite{song20203D} on 55 neuron specimens covering a depth range from -18 to 36 \si{um}. 
It shows that our approach outputs more structured sparse codes than competing methods owing to supervised learning with the soft-labels. Sparse codes of high quality facilitate depth detection via finding non-zero support centers more accurately. In contrast, Phase-Space~\cite{liu20153d,pegard2016compressive} and CSC~\cite{song20203D} tend to find more small non-zero elements, which may interfere with subsequent depth detection, as shown in Figure~\ref{Fig:SparseCodes}. Further results in Figure~\ref{Fig:Compare3Methods_Cell} confirms that our approach outperforms the competing methods and gives the best performance with least errors on x, y, z position detection. Furthermore, CISTA-net is more than 10000$\times$ faster than the competing methods in terms of average time for processing a single EPI with the same CPU, as shown in Table~\ref{Tab:TimeCost}.
CISTA-net computational efficiency can enable fast source localization during LFM imaging of living brain tissues, providing information to guide and modify the experimental protocol.


\begin{table}[t]
	\centering
	\small
	\caption{Average time cost of different methods.}
	\begin{tabular}{c|c|c|c}
		\hline \hline 
		& phase-space~\cite{liu20153d,pegard2016compressive} & CSC~\cite{song20203D} & CISTA-net \\ 
		\hline
		time cost  & 3.59 s & 3.61 s & 0.216 $\times 10^{-3}$ s \\
		\hline \hline
	\end{tabular} 
	\label{Tab:TimeCost}
\end{table}


\vspace{-0.2cm}

\section{Conclusion}
\label{sec:Conclusion}

\vspace{-0.2cm}

We proposed a model-inspired deep network to perform 3D localization using light-field microscopy images. The network architecture is designed systematically by unrolling the convolutional ISTA method used to solve convolutional sparse coding problems. The network parameters are learned from the training dataset. In this way, the designed network naturally inherits domain knowledge and also learns new features from the data, demonstrating better interpretability and higher speed. 
The proposed methodology may provide inspiration on how to take better advantage of the unrolling idea to design deep networks with advanced attributes.


\clearpage

\bibliographystyle{IEEEtran} 
\bibliography{mybib_LFM.bib}

\clearpage

\end{document}